\documentclass[runningheads, 10pt]{llncs}

\usepackage{graphicx}
\usepackage[most]{tcolorbox}
\usepackage[font=scriptsize,labelfont={scriptsize,bf}]{subfig}
\usepackage[normalem]{ulem}
\usepackage{rotating}
\usepackage{tabularray}
\usepackage{booktabs}
\usepackage{tabularx}
\usepackage{multirow}
\usepackage{pgfplots}
\pgfplotsset{compat=1.18}
\usepackage{amstext}  
\usepackage{enumitem}
\usepackage{xspace}
\usepackage{amssymb}
\usepackage[overload]{empheq}

\usepackage[font=scriptsize]{caption}
\usepackage{graphbox}
\usepackage[colorlinks=true,linkcolor=blue,breaklinks=True,citecolor=brown,urlcolor=blue]{hyperref}
\usepackage[dvipsnames]{xcolor}
\usepackage{soul}
\usepackage{xurl}
\usepackage{pifont}

\setlist{noitemsep, topsep=0pt,leftmargin=*}

\newcommand{\cmark}{{\color{ForestGreen}{\ding{51}}\xspace}}
\newcommand{\xmark}{{\color{RubineRed}{\ding{55}}\xspace}}

\newcolumntype{?}{!{\vrule width 1.5pt}}

\newcommand{\textbox}[1]{
    \noindent\fbox{%
        \parbox{0.97\columnwidth}{%
            {#1}
        }%
    }
}

\newtcolorbox{cooltextbox}[1][]{%
    colback=black!5,
    colframe=black!5,
    notitle,
    sharp corners,
    borderline west={0pt}{0pt}{red!80!black},
    enhanced,
    breakable,
    left=0pt,
    right=0pt,
    top=0pt,
    bottom=0pt
    }

\newcommand\smamath[1]{{\small $#1$}}

\newcommand\scmath[1]{{\scriptsize $#1$}}

\newcommand\purple[1]{%
  \bgroup
  \hskip0pt\color{purple!50!blue}%
  #1%
  \egroup
}

\begin{document}
\title{Can SOC Operators Explain their Decisions while Triaging Alarms? A Real-World Study}

\titlerunning{Can SOC Operators Explain their Decisions while Triaging Alarms?}

\author{Jessica Moosmann\inst{1} \and Irdin Pekaric\inst{1} \and Giovanni Apruzzese\inst{2}\inst{1}}

\institute{University of Liechtenstein, Vaduz, Liechtenstein \and Reykjavik University, Reykjavik, Iceland}

\maketitle 

\begin{abstract}

Security Operations Centers (SOCs) are pivotal in modern enterprises. Tasked to monitor complex network environments constantly under attack, SOCs can be active 24/7 and can include hundreds of operators supported by state-of-the-art technologies. Abundant research has studied the internal processes of SOCs, highlighting their pros and cons, as well as the challenges faced by SOC analysts---such as dealing with the overwhelming number of false alarms triggered by automated security mechanisms. In this context, we wonder: given that ``someone'' must triage the alarms, and that such triaging must be grounded on established knowledge or evidence-based reasoning, \textit{can SOC employees justify why a certain decision was taken while triaging alarms?}
Answering such a research question (RQ) can better guide future efforts. 

We hence tackle this RQs. First, via a systematic literature review across 257 research documents, we provide evidence that such RQ received limited attention so far. Then, we partner-up with a real-world SOC and carry out a field study (n=12) with SOC employees. We show them real alarms raised in their SOC, and inquire whether such alarms are indicative of true security problems or not. Then, we ask to explain their decision. We found that while most analysts were able to separate ``true from false'' alarms (the decision was correct in 83\% of the cases), a correct justification was hardly provided (only 39\% of the provided explanations reflected the actual root cause).
Ultimately, our results highlight the need for decision-support systems that help SOC analysts not only make the right call---but also understand and articulate \textit{why} it is right.

\end{abstract}

\section{Introduction}
\label{sec:introduction}

\noindent
Organizations worldwide are constantly targeted by cyberattacks~\cite{wef2025global,checkpoint2025global,deepstrike2025top}. To address this problem at scale, Security Operation Centers (SOC) are a fundamental asset to businesses of any size~\cite{clearnetwork2025soc,sans2025soc}. Mostly operating 24/7 and typically having 2--10 staff members~\cite{sans2025soc}, SOC protect the network perimeter of various companies by preventing incoming threats, detecting intrusions, as well as implementing and enacting recovery plans~\cite{sans2025soc}. To this purpose, SOC leverage state-of-the-art tools, such as system information and event management (SIEM) platforms, which are kept up-to-date with the most recent security feeds~\cite{sans2025soc}. 

\begin{figure}[!t]
    \centering
    \includegraphics[width=1\columnwidth]{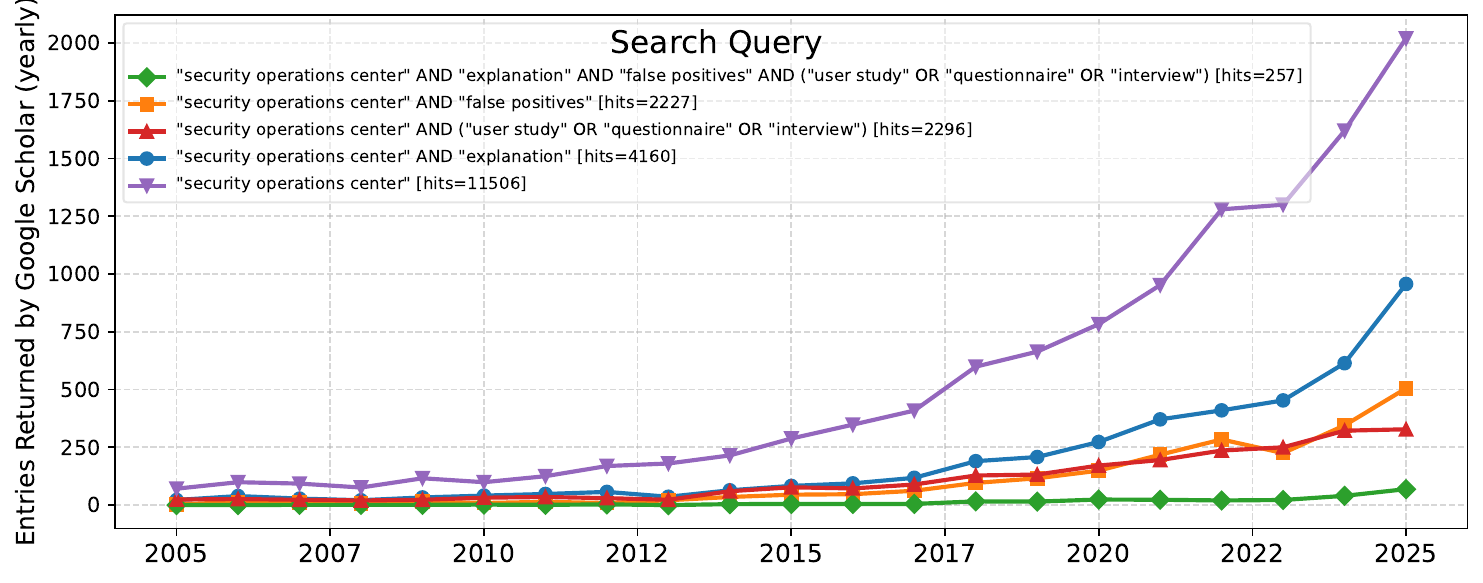}
    \vspace{-8mm}
    \caption{\textbf{Research interest in Security Operations Centers (SOC).} Results returned by issuing specific queries in Google Scholar. For ``security operation centers'', we considered variants (e.g., the complete term was ``(`security operation center' OR `security operations center' OR `security operation centre' OR `security operations centre'); for ``false positives'' we considered: ``(`false alarm' OR `false positive' OR `false alert')''; for ``user study'', we considered ``(`user survey' OR `user study')''; for ``explanation'', we consider ``(`explainability' OR `explainable' OR `explain' OR `explanation')''.} 
    \label{fig:trends}
    \vspace{-6mm}
\end{figure}

The criticality of SOC in the current information-technology (IT) landscape is established~\cite{sans2025soc}. As such, lots of scientific research focused on SOC, proposing methods to improve their workflows~\cite{vermeer2023alert}, enhancing deployed systems~\cite{van2022deepcase,sopan2018building}, analysing pain points~\cite{alahmadi202299}, or exposing their weaknesses~\cite{reeves2025s}. To portray the historical ``research interest'' towards SOCs, we show in Fig.~\ref{fig:trends} the number of results returned (yearly) by Google Scholar by issuing various SOC-related queries in the last 20 years. Accordingly, over 11k scientific documents have tackled the domain of SOCs. Recurrent themes involve, e.g., the problem of \textit{false positives}, triggered by the overwhelming amounts of data processed by various devices which must be analysed and upon which high-stakes decisions must be taken~\cite{mink2023everybody}. Such processes lead to the well-known \textit{alert fatigue} and decrease the efficiency of SOC employees~\cite{tariq2025alert}. Nevertheless, SOC analysts are also expected to \textit{explain} why a certain decision was taken---especially when they need to justify, to their customers, why an incorrect decision was made~\cite{apruzzese2023sok}.

In this context, and also given that SOC-related infrastructures are recognized also by legal bodies, industry standards, and institutional boards~\cite{nccs2024,nisdirective2016,iso27001:2022}, we took a step back, and one research question (RQ1) surfaced. \purple{\textit{Can SOC analysts explain their decisions while triaging alarms?}} Indeed, modern environments generate alarms, not all of them being indicative of security-noteworthy events~\cite{meschini2024case}; for instance, Yang et al.~\cite{yang2024true} claim that most alarms are due to so-called \textit{benign triggers}, deriving from apparently malicious behaviors but stemming from benign causes (e.g., a scheduled vulnerability scan launched by a security appliance). We hypothesize that some of these alarms can be trivial to resolve, especially if appropriate context is given. But what if the alarms are more challenging to triage? In these cases, and given that a decision must be made (e.g., either to invest resources in troubleshooting the problem---potentially leading to time waste; or ignore the problem---and potentially causing a security breach), can the operators provide a plausible explanation for their choices?

We argue that RQ1 has plenty of practical implications. For instance, if the answer is a clear ``yes'', then it means that existing infrastructures (encompassing both IT systems as well as documentation and management) provide good support to SOC analysts. Conversely, if the answer is ``not so much'', then it would indicate that there is a need for better decision-support systems~\cite{khayat2025empowering}. Regardless, tackling RQ1 also simultaneously enables one to address another research question (RQ2): \purple{\textit{Are certain alarms harder to triage than others?}} Indeed, investigating RQ1 requires to show various alarms to SOC operators and record {\footnotesize \textit{(i)}}~the decision and {\footnotesize \textit{(ii)}}~the explanation. Although the 2019 study by Kokulu et al.~\cite{kokulu2019matched} argues that false positives do not have a significant impact on the operation of the considered SOC, we believe that exploring RQ1 and RQ2 is instrumental to guide future research: if certain ``false alerts'' are systematically easier to triage, then specific rules can be defined so as to avoid burdening analysts---whereas ad-hoc tools would be needed to handle more complex cases.

\vspace{1mm}

\noindent
\textsc{\textbf{Contributions.}} In this work, we conduct a case study in a real-world SOC to investigate RQ1 and RQ2. Let us summarize our major contributions:
\begin{itemize}
    \item As a first step, we examine if RQ1 had been previously explored. Through a systematic literature review across 257 related documents, we found only one peer-reviewed work~\cite{kersten2024security} that can be said to be truly related to RQ1. However, the research in~\cite{kersten2024security} did not specifically focus on RQ1, since the respective user study (n=4) was meant to validate the output of a research prototype. 
    \item Then, to shed more light, we find an agreement with a SOC in Europe and carry out a field exercise with 12 full-time analysts (§\ref{sec:method}). Our study encompasses over 30k events distributed across six cases of different triaging difficulty. For each case, participants must provide a decision (true or false alarm) and justify their choice via open-text answers. The decision was correct in 83\% of the answers, but 61\% of the explanations were incorrect or imprecise (§\ref{sec:results}).
    \item Qualitative analyses based on a pre- and post-study questionnaire revealed that expertise or self-confidence do not reflect a participant's explanation accuracy. Moreover, some cases were perceived as systematically harder to triage, with participants commenting that there was ``missing information'' (even though they were provided with all the necessary tooling). 
\end{itemize}
We do not seek to generalize our results to all SOCs, but our findings can be instrumental for future work (§\ref{sec:discussion}). We provide additional details in our repo~\cite{repository}. 
\section{Background and Motivation}
\label{sec:related}
\noindent
We outline fundamental concepts of SOCs (§\ref{ssec:soc}), and then discuss our literature review (§\ref{ssec:slr}), which underpins the research gap tackled by our work (§\ref{ssec:gap}).

\subsection{Security Operations Centers (SOC)}
\label{ssec:soc}
\noindent
Security Operations Centers have been extensively covered in prior research (refer to Fig.~\ref{fig:trends}). Recent summaries and reviews, such as~\cite{tariq2025alert,saha2025expert,khayat2025empowering}, provide a comprehensive overview of the current state of SOCs. In what follows, we pinpoint the elements that are most necessary to appreciate our contributions.

\begin{figure}[!h]
    \centering
    \vspace{-6mm}
    \includegraphics[width=0.6\columnwidth]{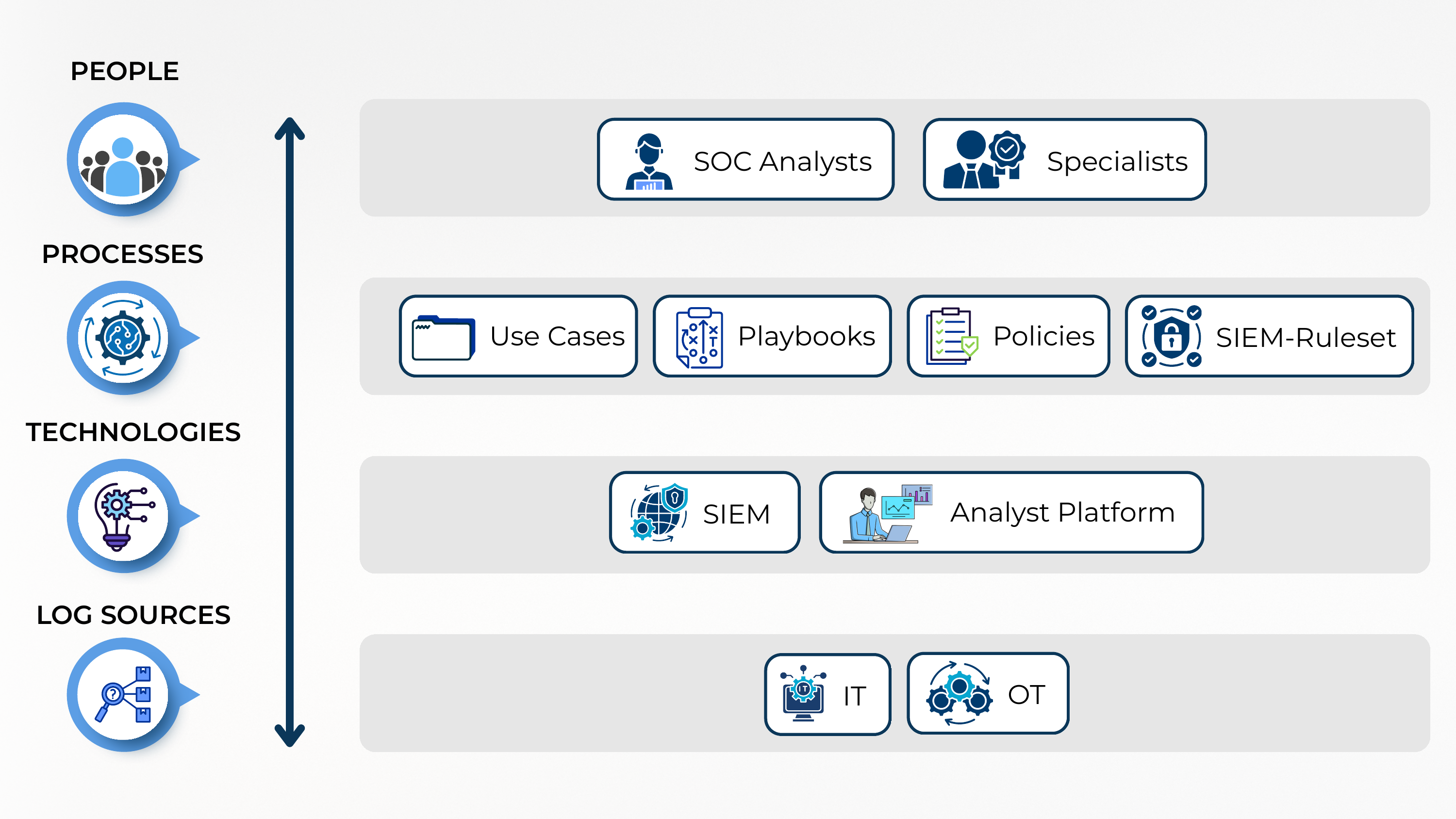}
    \vspace{-6mm}
    \caption{Typical SOC architecture (IT-Information Technology, OT-Operational Technology).} 
    \label{fig:sok_arch}
    \vspace{-9mm}
\end{figure}

A SOC is a central unit within a company whose main task is to (continuously) monitor their Information and Communication Technology environment~\cite{zimmerman2014}. It is common~\cite{zimmerman2014,alahmadi202299} to distinguish three key components that, if properly orchestrated, lead to an effective SOC: \textit{people}, \textit{processes}, and \textit{technologies}. We provide in Fig.~\ref{fig:sok_arch} a sketched architecture of a typical SOC (supported also by the analysts of our considered SOC). SOCs teams typically encompass 2--10 \textit{people}~\cite{sans2025soc}: while (security) analysts are those responsible for handling security incidents, other specialized personnel (e.g., engineers) are also crucial to ensure a smooth workflow~\cite{kokulu2019matched,stevens2022ready}. \textit{Processes} serve to establish the necessary boundaries that define how \textit{people} interact with \textit{technologies}. For instance, instructions as to how to deal with alarms, what activities are or not allowed by certain hosts, or who to contact in case of a breach---all such information can be found in process-related documents (typically known as \textit{playbooks}~\cite{stevens2022ready}). \textit{Technologies} encompass all the technical infrastructure that embeds the core functionality of the SOC. The most popular tools include SIEMs~\cite{yang2024true}, which typically are developed by well-known security vendors (e.g., Splunk~\cite{splunk}); however, a SOC can also rely on ad-hoc and custom software. Such software is designed to receive data inputs from a variety of \textit{Log Sources}. Traditionally, such log sources encompass IT devices, such as firewalls or routers; however, in select SOCs, these log sources can also include ``Operational Technology'' (OT) devices, such as PLC or components typically deployed in industrial control systems~\cite{colelli2019securing,lipnicki2018future}.

Numerous long-standing problems affect SOC environments. For instance, false alarms---for which it is even hard to even provide an universally-recognised definition (e.g., the ``benign triggers'' envisioned by~\cite{yang2024true} are not considered false positives according to~\cite{alahmadi202299}; and regardless, Kokulu et al.~\cite{kokulu2019matched} argue that, after all, false positives do not have a significant impact on SOCs). Another closely related issue is that of \textit{explainability}~\cite{mink2023everybody,apruzzese2023sok}. SOC environments are becoming increasingly more complex, and analysts must make more decisions which, in turn, should account for more data. As such, analysts are on the lookout for tools that ``explain'' the root cause of a certain event (or, alternatively, ``why an alarm was raised''). Such additional input can hence be used to make an informed decision in the presence of overwhelming amounts of datapoints~\cite{alahmadi202299,nadeem2023sok}. Otherwise, as previous studies indicated~\cite{cho2020,ofte2024,reeves2023}, analysts may rely on intuition or gut feelings, which can cloud their judgment and lead to irrational decisions.

\subsection{Systematic Literature Review}
\label{ssec:slr}
\noindent
We began our study with a broad research question (RQ0): \purple{\textit{has prior work carried out user studies in which SOC practitioners were asked to triage alarms and explain their decisions?}} To answer  RQ0, we carry out a systematic literature review (SLR), following established PRISMA guidelines~\cite{moher2015preferred}. A depiction of our methodology is in Fig.~\ref{fig:slr} (more details are reported in the Appendix~\ref{sapp:slr}). 
\footnote{We carried out a preliminary investigation in Dec. 2024; we repeated this process in Dec. 2025 to ensure that, at submission time, our answer is still valid; and further validated our findings by double-checking~\cite{franklin2001reliability} our results a third time in Feb. 2026.}

\textbf{Paper Collection.} We collected works via Google Scholar. We devised a search query revolving around four keyword groups: ``SOC'', ``explainability'', ``user study'', and ``false positives''. Intuitively, a paper that addressed RQ0 (implicitly or explicitly) should mention these terms at least once (refer to the caption of Fig.~\ref{fig:trends} for the exact queries). We filter the results between 2005--2025 (the earliest document appeared in 2007~\cite{goodall2007defending}). Overall, we obtain 257 papers.

\textbf{Screening.} Next, we remove results not relevant for RQ0. We remove all non-peer-reviewed works (e.g., preprints~\cite{singh2025llms} or theses/books~\cite{goodall2007defending,chung2023interactive}). This way, we obtained 74 papers. Then, we manually checked the content of these papers, seeking to identify papers that \textit{carried out user studies}. 19 papers only mentioned terms related to user studies as generic recommendations (e.g.,~\cite{teuwen2025ruling}) or to refer to prior work (e.g.,~\cite{tariq2025alert}). However, while inspecting such works, we found one paper that is very relevant for RQ0, i.e.,~\cite{eriksson2022towards} which we added to our SLR.\footnote{This work was not captured by our query because~\cite{eriksson2022towards} it mentions ``explainable alerts'' and not ``false alerts''; all works relevant to RQ0 should mention ``false alerts/alarms/positives'', but to avoid missing these cases we inspected also the references. Specifically, we found~\cite{eriksson2022towards} while reviewing the recent survey by Tariq et al.~\cite{tariq2025alert}.}

\textbf{Analysis.} We manually analysed our set of 55 papers carrying out user studies. We sought to answer two questions: {\footnotesize \textit{(i)}}~``what type of user study is carried out?''
{\footnotesize \textit{(ii)}}~``are the participants SOC practitioners?''.
For the first question, we found that 34 papers carried out interviews, whereas 9 used asynchronous surveys/questionnaire; four papers carried out a mixed-method study~\cite{saha2025expert,stevens2022ready,kersten2024security,ulmer2019netcapvis}, and seven carried out a field-study exercise~\cite{zhong2017studying,zhong2018learning,sundaramurthy2014tale,reeves2025s,jansen2024comparative,liu2021faixid,araujo2025enhancing}. 
For the second question, only 14 papers made it explicit that the participants were \textit{exclusively} SOC professionals: other works consider generic employees~\cite{burda2024protect}, or pooled SOC analysts with non-SOC members (e.g.,~\cite{naseer2023moving}), or a broad ``security specialists/professionals/researchers'' (e.g.,~\cite{hagen2025human,mink2023everybody}). Nonetheless, some papers clearly did not cover SOC environments (e.g.,~\cite{roul2025incident}): among our 55 papers, we could only map 39 to SOC-specific contexts (e.g., triaging alarms while accounting for real-world constraints). 
For transparency, we report the full results of our SLR in~\cite{repository}.

\begin{figure}[t]
    \centering
    \includegraphics[width=\linewidth]{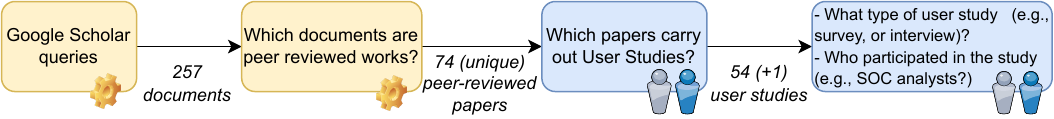}
    \vspace{-7mm}
    \caption{Overview of the methodology of our SLR. The qualitative analysis was done by two authors.}
    \label{fig:slr}
    \vspace{-5mm}
\end{figure}

\subsection{Related Work and Research Gap}
\label{ssec:gap}
\noindent
To fully answer RQ0, we assessed the papers found through our SLR once more, asking ourselves: ``is the paper about explaining the decisions made by analysts?'' In what follows, we briefly discuss our results, highlighting the works that are closest to our purpose---and which also support our overall research.

The authors of~\cite{liu2021faixid} carry out a proof-of-concept evaluation of a framework meant to study the explainability of AI-driven methods: this is done via a survey across 7 participants who should serve as a proxy of ``potential cybersecurity analysts''. However, aside from the unclear background of participants, the assessment is based on research prototypes, and does not reflect an operational SOC environment.
The seminal work by~\cite{alahmadi202299}, followed by other recent works from top-tier venues~\cite{saha2025expert,mink2023everybody}, contributed with the results of 15--21 interviews with SOC (or security-focused) professionals. However, while these works serve to highlight common operational pain-points, the findings are based on the perceptions of the interviewees---in other words, there is no actual assessment of the practitioners' ability to explain their decisions (or triage false alarms). 
The research in~\cite{eriksson2022towards} involves a user study, done with 10 SOC analysts, focused on assessing the quality of \textit{eXplainable AI} techniques in a SOC context. Despite being conceptually similar to our goal (i.e., assessing explanations), there is a fundamental difference because the study in~\cite{eriksson2022towards} focuses on assessing software-generated explanations, whereas we focus on human-generated explanations.

Three very close works are~\cite{kurogome2019eiger}, and~\cite{zhong2017studying,zhong2018learning}. The former~\cite{kurogome2019eiger} inquires 15 analysts within the same SOC about the perception of the output of the solution proposed in the corresponding paper, asking questions such as ``do you think you understood what this IOC means?'' which could be answered with a 5-point Likert scale: such an instrument, again, mostly reflects the perception of the participant. The authors of~\cite{zhong2017studying,zhong2018learning} conduct a full-fledged simulation (across ``30 professional analysts from Army Research Lab'') sought to measure the participants ability to triage alarms. However, the data used for this experiment was not that of the SOC wherein the participants worked, but was generated through a simulation which occurred years before the study took place; moreover, there is no assessment of the participants' ability to explain their decisions. 

The work most aligned with RQ0 is the paper by Kersten et al.~\cite{kersten2024security}. They carry out a study with four SOC practitioners asking which alerts are ``interesting'' (which can be a way to gauge if the practitioner considers it as a true/false alert) and to adopt a (recorded) think-aloud approach (which slows-down the analysis~\cite{pike2014measuring}). Then, participants motivated their choice via pre-defined answers. This is the only work we found that can be used to answer positively to RQ0.

\vspace{-1mm}

\begin{cooltextbox}
\textsc{\textbf{Answer to RQ0.}} According to our SLR,\footnote{We also applied the snowball method~\cite{wohlin2014guidelines}, checking for papers cited, or cited by, the aforementioned works. Despite finding other relevant papers (e.g.,~\cite{garneau2016results,thomson2024too}), we were unable to find other works that could support a positive answer to RQ0.} only one work~\cite{kersten2024security} could be said to have investigated whether real \textit{SOC practitioners} can correctly \textit{explain the decisions} they make. However, this study~\cite{kersten2024security} was conducted to assess a proposed tool, considered only 4 SOC analysts, and the justifications had to be provided via think-aloud mechanism (which increases the cognitive load and influence the analysis~\cite{pike2014measuring}) and
pre-defined answers (which may lead to bias~\cite{connor2019comparing}). 
\end{cooltextbox}

\vspace{-1mm}

Our work builds upon that of~\cite{kersten2024security}. Specifically, we will examine if employees of a single SOC, when faced with high-stakes decisions such as whether to consider an alarm as a false or true positive, can {\footnotesize \textit{(i)}}~make the correct choice, and {\footnotesize \textit{(ii)}}~justify their choice via open-text answers~\cite{choi2004catalog}. Importantly, we acknowledge that focusing on a single SOC prevents deriving generalizable conclusions---which is why we do not make such a claim. However, we believe that our contributions are useful to {\footnotesize \textit{(a)}}~inspire future work to carry out similar studies; and {\footnotesize \textit{(b)}}~provide evidence of how well some SOC practitioners can handle their routine tasks---which entail interacting with state-of-the-art production~systems. 
\section{Methodology}
\label{sec:method}
\noindent
We describe the approach used to investigate how well SOC analysts triage alarms and justify of their decisions. We utilize a controlled field experiment, carried out in a real SOC, entailing structured questionnaires and qualitative analysis. We aim to obtain factual evidence of SOC analysts’ decisions in conditions mirroring operational environments, thereby enabling to answer our RQs. 

An important remark is that our study was devised in close collaboration with the examined SOC. This ensured that the setup aligned with the operational environment, thereby increasing the ecological validity of our study (We will discuss the implications of this peculiar characteristic of our study in §\ref{ssec:limitations})

\subsection{Study Design Overview} 
\label{ssec:study_design}

\noindent
Our methodology follows a mixed-methods design consisting of two components: the definition of a realistic SOC case environment (which we discuss in §\ref{ssec:technical_environment}); and a practical alarm-classification exercise, enriched with a pre- and post-exercise questionnaires (explained in detail in §\ref{ssec:technical_implementation}). Our approach hence resembles the one in~\cite{kersten2024security,kurogome2019eiger}, with the difference that our study does not revolve around validating a new ``tool/solution'', rather, our study is grounded in the operational workflow of a real-world SOC (from both a data and system perspective). 

At a high level, the considered SOC leverages state-of-the-art appliances, including a SIEM platform that aggregated reports from various log sources. From this environment, we derive a set of representative alarms (which we denote as \textit{cases}) that encode varying degrees of difficulty, contextual complexity, and ambiguity. Alarm selection was guided by three criteria:~diversity of underlying causes (e.g., benign processes, misconfigurations, scanning activity); variation in contextual dependencies; and mixed explainability challenges, ranging from trivial to essentially ambiguous alerts. This design allows us to isolate differences in analysts’ abilities to detect false alarms and articulate correct reasoning. Yet, given the space of use cases that can occur in a SOC, we anticipate that any selection (for the sake of our study) cannot ensure complete coverage.

For time-related reasons (further detailed in §\ref{ssec:technical_implementation}), we ultimately decided to include six \textit{cases} in our study, summarized in Table~\ref{tab:cases}. Such a selection was carried out qualitatively by two researchers: one with access to domain-specific operational knowledge of the SOC, and one independent researcher, who acted as a validator for the selection. Collectively, our \textit{cases} encompass over 30k events which triggered 24 alarms in total. Recall that, for RQ2, we want to see if certain alarms are harder to triage: therefore, choosing \textit{cases} that -- according to our own judgment -- are more complex should serve to question such an hypothesis, which we can test both by measuring the rate of correct answers, but also via the follow-up questionnaire. This is why some \textit{cases} in Table~\ref{tab:cases} are marked as ``trivial'', meaning that a SOC analyst should quickly infer that the alarm is not indicative of a security-noteworthy event. On the contrary, other \textit{cases} are marked as ``relevant'' because making such a decision requires carrying out additional analyses to obtain more contextual information.

After defining the \textit{cases} (further described in our repository~\cite{repository}) and after recruiting the participants (we report some demographic details in Appendix~\ref{sapp:demographics}), we administered the user study (in April 2025) and analysed the results.

\begin{table}[!t]
\tiny
\centering
\caption{\textbf{Summary of the Six Cases of Alerts included in our Study.} Brief description: \textit{Case-1}: Scheduled vulnerability scan triggered alarms. \textit{Case-2}: Firewall allows specific developer traffic. \textit{Case-3}: DNS sinkhole detects C\&C traffic; connection blocked. \textit{Case-4}: HTTP redirect to \texttt{msftconnecttest} misclassified as suspicious. \textit{Case-5}; WiFi client triggered alert due to network instability. \textit{Case-6}: Endpoint protection quarantined infected file. (More details in our repository~\cite{repository})}
\label{tab:cases}
\resizebox{\columnwidth}{!}{
\begin{tabular}{p{1.5cm}| p{1.3cm}| p{0.7cm}| p{0.9cm}| p{1.9cm}| p{2.3cm}| p{1.2cm}| p{2cm}}
\toprule

\textbf{Case} &
\textbf{Category} &
\textbf{Alerts} &
\textbf{Logged Events} &
\textbf{Alert Type} &
\textbf{Key \break Observables} &
\textbf{Source} &
\textbf{Action / \break Indicator} \\
\hline

\textbf{1: Network scan by internal system} &
False Positive - trivial (with context) / relevant (without) &
15 &
2,034 over 4h &
Port scanning / suspicious network activity &
IP source, IP destination, hit count, firewall action &
Internal IP (Nessus server) &
Accept; system role identified in asset database \\
\hline

\textbf{2: Developer activity triggering alerts} &
False Positive - trivial &
4 &
1,340 over 48h &
Outbound connections from internal IPs &
IP source, IP destination, hit count, protocol, firewall action, category, properties &
Internal developer-segment IPs &
Accept; triggered by developer-specific firewall rule \\
\hline

\textbf{3: Outbound C2 communication} &
True Positive &
1 &
2 over 10h &
Outbound connection to Command \& Control server &
IP source, IP destination, hit count, firewall action, malware action &
Internal host &
Prevent; DNS trap and malware action confirm C2 activity \\
\hline

\textbf{4: Windows connectivity check} &
False Positive - trivial &
2 &
31,518 in 24h &
Outbound HTTP/S traffic to connectivity-check URL &
IP source, IP destination, hit count, protocol, firewall action &
Internal load balancer &
Pass; traffic to \texttt{msftconnecttest} incl.\ HTTP redirect \\
\hline

\textbf{5: Internal retransmission classified as malicious} &
False Positive - relevant &
1 &
2 in 2h &
Internal retransmission pattern flagged as malicious &
IP source, IP destination, hit count, protocol number, detected host, domain, action, reason &
Internal dynamically assigned WiFi IP &
Block; triggered because IP source belongs to WiFi pool \\
\hline

\textbf{6: Malicious PowerShell script} &
True Positive &
1 &
2 in 20h &
Endpoint protection alert: suspicious PowerShell script &
Hit count, detected host, host, domain, user, label, reason &
Endpoint (self-reported) &
File blocked \& quarantined; virus label, affected file, external-account username \\
\bottomrule
\end{tabular}
}
\vspace{-7mm}
\end{table}

\subsection{Technical Environment (SOC Description)} 
\label{ssec:technical_environment}

\noindent
We conducted our study in a SOC operating in the DACH (German-Austrian-Swiss) area; for confidentiality, we cannot provide extensive details. Companies in this geographical location are constantly under cyberattacks~\cite{bsi2024,ey2024,sci2024}, meaning that our considered environment can serve as a proxy for SOCs in other areas.

The SOC's architecture resembles that described in §\ref{ssec:soc}. The SOC monitors a hybrid IT/OT environment that includes heterogeneous network segments, Windows and Linux endpoints, perimeter firewalls from multiple vendors, industrial controllers and several business-critical applications. All these systems continuously generate audit, network, and security telemetry. This data is sent to a central SIEM platform via standardized interfaces (e.g., syslogs). 

The SIEM constitutes the analysts’ primary work environment. It runs the production correlation pipeline that consists of data normalization, enrichment, rule-based matching and temporal correlation. We did not add any experimental rules. All the alarms that were used in the exercise were triggered using the standard rule set, which is operated in daily business. This ensures that the applied \textit{cases} reflected typical data quality, data volume, enrichment depth, and alert semantics that is encountered during regular operations (unlike the synthetic setup in~\cite{zhong2017studying,zhong2018learning}, using data from elsewhere).

The SOC uses a layered tool stack that consists of {\footnotesize \textit{(i)}}~the SIEM for raw-data inspection and correlation logic, {\footnotesize \textit{(ii)}}~a dedicated case-management platform where alarms are triaged and documented, and {\footnotesize \textit{(iii)}}~an internal asset database that provides contextual information such as device roles, network segments, ownership and previous activities. During the study, participants received the same access rights and data views that they use in routine operations. We did this without introducing any additional filters, hints or experimental artifacts (which, perhaps unconsciously, could influence participants' responses~\cite{kersten2024security,kurogome2019eiger}).

Our technical environment thus ensured that participants’ decisions are shaped by the same constraints, information structures and tooling conditions of their routine SOC environment. The six addressed \textit{cases} (in Table~\ref{tab:cases}) were extracted from this ecosystem without modifying event semantics, field names, or rule logic. Moreover, all \textit{cases} were derived from alerts that had occurred in production and had been processed through the standard normalization, enrichment, and correlation pipeline. Moreover, we considered \textit{cases} that {\footnotesize \textit{(i)}}~occurred weeks before the field study, thereby ensuring that participants would not remember them; but which {\footnotesize \textit{(ii)}}~were still affected by the same infrastructure, thereby ensuring that the corresponding ground truth did not change.

Raw log entries associated with each alert were exported together with the correlated event groups that the SIEM generated at the time of detection. During preparation, sensitive fields (e.g., internal hostnames, user identifiers, customer-related data) were anonynimised according to the company’s data-protection guidelines. We preserved the structural characteristics required for triaging, including timestamps, alert metadata, correlation fields and observable attributes.

Each \textit{case} was validated against the operational incident documentation and the asset database to confirm its factual ground truth. This ensured that both true positives and false positives reflect real underlying causes and not synthetic constructs. Importantly, sanitization process did not alter the semantic content of the logs or introduce potential artifacts that could non purposely guide analysts. The final \textit{case} sets were prepared in a way that maintained the full analytical complexity of the original alerts. Additionally, they met all internal confidentiality and data-protection requirements.

\subsection{Study Implementation and Ethics} 
\label{ssec:technical_implementation}

\textbf{Participant Recruitment.} In April 2025, we recruited 12 full-time security analysts that work in the same SOC environment. Participants were of various experience levels, ranging from junior to senior analysts. Participation was completely voluntary, and there was no identifiable information collected. The study was carried out during the participants' regular working hours. We quantified the total time that each participant would spend for our study in roughly 3 hours, of which 2 were spent on the day of the exercise, and another hour would be spent for logistics and communication. To better reflect operational contexts and to minimize time waste, the study was done synchronously (on site) and in groups, but participants could not communicate to each other. This ensured that their decisions could be attributed to personal reasoning. Importantly, the employer company had been made aware of our study, and the overall design and implementation complied with the SOC's internal standards (confirmed by their management). We report in the Appendix~\ref{sapp:ethics} additional ethical considerations.

\vspace{1mm}

\noindent
\textbf{Pre-study phase.} 
The study begins with a \smamath{\approx}45 minutes preparatory session during which participants are explained the objectives and procedure of the study. This included an explanation of what to expect over the next two hours, and which systems are available for processing the tasks (e.g., the SIEM). Organizational framework conditions were also discussed, including the planned time structure, instructions on behavior during the study and how to deal with open questions. A compact introduction to the systems was given, in which the most important functions relevant to case processing were explained. Finally, the technical setup was reviewed and tested and any final questions were clarified. After this preparatory phase, the participants were given access to a preliminary questionnaire (entirely reported in our repository~\cite{repository}), through which we could record participants' previous experience, skills and assessments with regard to alarm analyses. The average time taken to complete the initial questionnaire was around 15 minutes. The aggregated results are provided in Appendix~\ref{sapp:demographics}. 

\vspace{1mm}

\noindent
\textbf{Practical Classification Exercise.} Participants were given a cheat sheet containing the key information to support the analysis. The duration of the exercise was set to 30 minutes: this served to infer how well each participant would perform in a limited timespan (roughly, 5 minutes per case). We report all cases in detail in Table~\ref{tab:cases}. For each case, participants performed two tasks: classification (they had to decide whether the case represents a true or false positive) and justification (they had to provide a free-text explanation that includes the rationale behind the classification). For the latter, and crucially, participants should have known how to properly justify each answer (both because of their prior expertise, but also because of the preliminary phase). Each case consisted of sanitized log data and contextual information extracted from the real SOC environment. The presentation mirrored the information flow that analysts experience during daily operations, including correlated events, timestamps, and relevant system attributes. We ensured that participants could access all information necessary to resolve a case (though usage of AI was prohibited, since it is also not part of the SOC workflow), and participants were reminded at regular intervals to move on to the next case if they were finding a case particularly challenging. Participants were not informed about the number of true/false positives. With the exception of one participant (who required 18 minutes), all participants took the entire 30 minutes to complete the exercise.

\vspace{1mm}
\noindent
\textbf{Post-study phase.} After the main exercise, participants were shown a survey asking open questions about the complexity of the cases, decision making, and time efficiency. The entire questionnaire is reported in our repository~\cite{repository}. 
The focus was on identifying influencing factors that have a positive or negative impact on the efficiency and accuracy of the analysis. Roughly 30 minutes were taken to complete this questionnaire. Finally, after reviewing the results, we communicated them to participants and asked for (unstructured) feedback.

\subsection{Qualitative Coding Approach}
\label{ssec:coding}
\noindent
We are not aware of prior studies with a scope similar to ours. Given the unstructured nature of the justifications, we adopted \textit{hybrid coding} approach, mostly reliant on inductive reasoning, but for which we followed high-level guidelines. 

\textbf{Coders.} Three reviewers (R1, R2, R3) participated in the qualitative assessment. R1 had access to contextual information provided by the participating SOC, enabling accurate interpretation of explanations; 
R2 is a security researcher with a decade of experience in this research domain, and who also has SOC work experience (in a different SOC); R3 is a security researcher with \smamath{>}5 years of experience in the operational security research domain. Importantly, R1 and R2 also contributed in the selection of the six cases considered in our study.

\textbf{Workflow.} After administering the study, the responses were first inspected by R1: for this preliminary assessment, R1 interacted with R2 to determine how to handle specific occurrences (e.g., ``what if a participant writes `no more time' in the open-text? Should it count as a `N/A' or as a wrong answer?''). Then, the coders used their own judgment to assign their codes to the responses. To avoid bias, R3 \textit{never interacted} with R1: R3 only used the detailed information on each case (reported in the supplementary material). With regard to R2, given that the study was done in German (the language of the SOC), machine-generated translations were provided to guide the assessment (whereas R1 and R3 are fluent in German). Once all coders assigned their codes, the opinions were compared, showing that coders were generally in agreement (there was consensus in 82\% of the codes). Discrepancies were resolved via synchronous and asynchronous communication, with R2 acting as middle point. Decisions were finalized in a final meeting between R1 and R2, during which some initial codes by R1 changed. Hence, even though R1 had the final call, the collective contribution of all coders was essential to maximise interpretative quality and derive accurate results. 

\textbf{Guidelines.} To assign codes, we established two rules: 
first, an answer was treated as ``N/A'' if it was left empty or if provided no form of technical assessment (e.g., ``no time'');
second, if the alert was classified incorrectly, then the explanation is never considered as correct (to avoid biasing our results).\footnote{We still inspected these answers, confirming our choice (explanations were not valid).} Then, each coder used their own judgment to assign codes. At a high-level, when analysing each response, we sought to answer two questions: {\scriptsize \textit{(a)}}~``does the text show understanding of the reason why the alarm was classified in a specific way?'' and {\scriptsize \textit{(b)}}~``is the explanation of use to another security analyst?''. A positive answer to either of these two questions would typically lead to a valid explanation (\cmark{}). This is because we are not merely interested in determining if, e.g., the answer is ``just'' correct,\footnote{Recall that our goal is determining if SOC analyst can motivate their decisions---which implies that the explanation should explain why a given alert was deemed as an FP or TP. For instance, for TPs, an explanation stating ``this is suspicious'' is not acceptable because it does not have any practical utility; whereas, for FPs, an explanation stating ``everything is normal'' would also be deemed as invalid, since it is obvious that ``everything is normal'' if an alert is considered an FP.} rather we want to see if the analyst is aware of what is going on in the monitored environment. We note that participants were explicitly told to be as specific as possible when motivating their decisions.

\section{Results}
\label{sec:results}

\noindent
We first examine the relationship between analysts’ classification performance and their ability to provide accurate and technically grounded justifications~(§\ref{ssec:rq1}). Finally, we investigate differences in the difficulty of evaluating false positives~(§\ref{ssec:rq2}).

\subsection{Accuracy and Quality of Analysts’ Explanations (RQ1)} 
\label{ssec:rq1}

We provide the results on the relationship between participants’ classification decisions and corresponding justifications. Classification represents the correctness of the final decision (true/false positive), while justification refers to providing a valid explanation that describes the actual technical cause of the alert. We recall (§\ref{ssec:technical_implementation}) that participants were given clear instructions (and should also have known) about how to justify their choices.

\vspace{1mm}

\begin{table}[!t]
\centering
\caption{Distribution of classification-explanation result types across all participants.}
\label{tab:summary}
\resizebox{0.4\columnwidth}{!}{
\begin{tabular}{l|c|c|c}
\textbf{Classification} & \textbf{Explanation} & \textbf{n (72)} & \textbf{\%} \\
\hline
Correct & Correct & 22 & 31\% \\ \hline
Correct & Incorrect / Vague & 35 & 48\% \\ \hline
Correct & Missing & 3 & 4\% \\ \hline 
False / Missing & -- & 12 & 17\% \\
\bottomrule
\end{tabular}
}
\vspace{-7mm}

\end{table} 

\noindent
\textbf{Overall Performance Patterns.} Across all six cases and 12 participants (72 total decisions), we identified four distinct outcome types. As shown in Table~\ref{tab:summary}, 22 decisions (31\%) included both a correct classification and a correct explanation; 35 decisions (48\%), represented correct classification but incorrect or incomplete explanation, whereas 3 (4\%) decisions were correct but no explanation was provided. Finally, the remaining 12 (17\%) were misclassified or skipped.
These findings suggest a misalignment between recognizing the correct outcome and being able to justify that outcome. Even though 60 (out of 72, i.e., 83\%) cases were classified correctly, an appropriate explanation was given only in 37\% of these cases (i.e., 22 out of 60).
Based on this result, we argue that analysts’ judgments are driven by intuition, experience or heuristics, and not by explicit reasoning---or, if so, then analysts cannot properly convey such a reasoning.

\vspace{1mm}

\noindent
\textbf{Fine-grained Analysis.} We report in Table~\ref{tab:participant_results} the fine-grained distribution of classification (``C'') and explanation (``E'') across all cases and participants; \cmark{} and \xmark{} denote correct and incorrect answers, whereas $\square$ denotes cases with no answer. Let us discuss some interesting findings related to specific cases. 

\begin{itemize}
    \item \textit{Case~1}, which represents a scheduled Nessus vulnerability scan, was the only case for which all 12 participants produced correct classifications. In addition, eight participants provided a correct explanation by drawing on information from the asset database and the scanner’s unique identifier. The high explanation correctness rate shows that analysts can provide appropriate reasoning when alerts exhibit a single technical artifact (all coders agreed on this).

    \item \textit{Case~2} highlights the main difficulty that strongly relates to RQ1. Regardless that 11 participants classified the case correctly, only one provided an appropriate explanation. Most answers did not express understanding of the reason why the case was a false alert. 
    We posit that this shows that humans make correct guesses because something feels normal (some participants explicitly confirmed this), but without being able to motivate the overarching reason. 

    \item \textit{Case-3} was correctly identified as a true positive by 9 participants, but only 5 these gave a valid explanation. In some instances, the explanation simply reported the name of the alert itself (i.e., ``Suspected Botnet'') or was a copy-paste of the SIEM's output, both of which being not appropriate explanations.

    \item \textit{Case~4} and \textit{Case~5} also demonstrate poor explanatory performance. \textit{Case~4}, which involved misclassified Windows connectivity checks, required that analysts identify that the large volume of HTTP redirects was benign. 9 participants recognized this, but only two correctly explained why. Many explanations mentioned traffic regularity but ignored the \texttt{msftconnecttest} domain, which was central to the ground truth\footnote{\texttt{msftconnecttest} is legitimate Windows Network Connectivity Status Indicator traffic, but in this case the host generating it should never have done so (a true alarm).}. In \textit{Case~5}, which involves dropped WiFi retransmits, explanations were even weaker. Regardless that nine participants arrived at the correct false positive classification, justifications frequently used irrelevant proxy events or false assumptions about user behavior.

    \item For \textit{Case-6}, an appropriate explanation was given only by 4 out of 10 participants that correctly guessed the alert was a true positive. Among the not satisfactory explanations, some participants stated ``the tool was recognised as malicious by the EDR'' which is true, but not informative because the same logic can apply for any alert raised by the security system. In contrast, valid explanations provided more insight about why the alert deserved further scrutiny (e.g., one stated ``it is a PowerShell script from Edge, so it is suspicious'', thereby expanding the information provided by the security system).
\end{itemize}
(We report in Appendix~\ref{sapp:examples} some examples of how we coded some responses)

\begin{table*}[!t]
\centering
\caption{Per-Participant Classification and Explanation Results Across All Cases}
\label{tab:participant_results}
\setlength{\tabcolsep}{4pt}
\resizebox{0.7\columnwidth}{!}{
\begin{tabular}{lcccccccccccc}
\toprule
& \multicolumn{2}{c}{Case 1} & \multicolumn{2}{c}{Case 2} & \multicolumn{2}{c}{Case 3} &
  \multicolumn{2}{c}{Case 4} & \multicolumn{2}{c}{Case 5} & \multicolumn{2}{c}{Case 6} \\
\cmidrule(lr){2-3} \cmidrule(lr){4-5} \cmidrule(lr){6-7}
\cmidrule(lr){8-9} \cmidrule(lr){10-11} \cmidrule(lr){12-13}
\textbf{Participant} &
\textbf{C} & \textbf{E} &
\textbf{C} & \textbf{E} &
\textbf{C} & \textbf{E} &
\textbf{C} & \textbf{E} &
\textbf{C} & \textbf{E} &
\textbf{C} & \textbf{E} \\
\midrule
P1  & \cmark & \cmark & \cmark & $\square$ & \cmark & \cmark & \cmark & \xmark & \cmark & \cmark & \xmark & \xmark \\
P2  & \cmark & \xmark & \cmark & \xmark & \cmark & \xmark & \cmark & \xmark & \cmark & \xmark & \cmark & \cmark \\
P3  & \cmark & \cmark & \cmark & \xmark & \xmark & \xmark & \xmark & \xmark & \xmark & \xmark & \cmark & $\square$ \\
P4  & \cmark & \xmark & \cmark & \xmark & \cmark & \cmark & \cmark & \xmark & \cmark & \xmark & \cmark & \cmark \\
P5  & \cmark & \cmark & \cmark & $\square$ & \xmark & \xmark & \cmark & \xmark & \cmark & \xmark & \cmark & \cmark \\
P6  & \cmark & \cmark & \cmark & \xmark & \cmark & \xmark & \cmark & \cmark & \cmark & \xmark & \cmark & \xmark \\
P7  & \cmark & \cmark & \cmark & \xmark & \cmark & \cmark & \xmark & \xmark & \cmark & \cmark & \cmark & \cmark \\
P8  & \cmark & \cmark & \cmark & \xmark & \cmark & \xmark & \xmark & \xmark & \xmark & \xmark & \cmark & \xmark \\
P9  & \cmark & \cmark & \cmark & \xmark & \cmark & \cmark & \cmark & \xmark & \cmark & \xmark & $\square$ & $\square$ \\
P10 & \cmark & \cmark & $\square$ & $\square$ & \cmark & \xmark & \cmark & \xmark & $\square$ & $\square$ & \cmark & \xmark \\
P11 & \cmark & \xmark & \cmark & \xmark & \xmark & \xmark & \cmark & \cmark & \cmark & \cmark & \cmark & \xmark \\
P12 & \cmark & \xmark & \cmark & \cmark & \cmark & \cmark & \cmark & \xmark & \cmark & \xmark & \cmark & \xmark \\

\midrule
\textbf{Correct} &
12 & 8 & 11 & 1 & 9 & 5 & 9 & 2 & 9 & 3 & 10 & 4 \\
\bottomrule
\end{tabular}
}
\vspace{-7mm}
\end{table*}

\vspace{1mm}

\noindent
\textbf{Individual Reasoning Patterns.} To better understand our results, we looked at the answers in the post-study questionnaire. We hypothesize that differences in explanation quality cannot be attributed solely to experience or technical background. Indeed three top-performing participants (P4, P6, P7) had different prior roles (in IT support, networks, software development). However, their performance cannot be attributed to their domain-specific expertise (which, in theory, also other participants had) but to their approach. These participants: {\footnotesize \textit{(i)}}~referenced multiple log fields; {\footnotesize \textit{(ii)}}~triangulated between SIEM data and the asset database; {\footnotesize \textit{(iii)}}~searched for related patterns rather than isolated events; and {\footnotesize \textit{(iv)}}~provided detailed reasoning (e.g., ``Hack tool in the name + Strange user $\rightarrow$ Pentest from \{redacted\}'' referring to Case-6).

In contrast, the three lowest-performing participants (P3, P8, P10) demonstrated very different behaviors. One participant submitted answers rapidly (18 minutes out of 30 available, denoting that time was not a factor), expressed high confidence, but did not look at the detailed fields within each alert, such as IP addresses, domains or log attributed. One explanation stated ``action was prevented, it is OK''. Another participant expressed persistent uncertainty and confusion: accordingly, this was due to difficulties in managing the cognitive load of the SIEM-like interface---despite such interface being the one used in the SOC. 

We also found that participants expressing high confidence not necessarily provide highly-accurate explanations. This was also the case for the experience level. For example, a participant with more than ten years of IT experience produced one of the weakest reasoning. On the other hand, a participant with low self-rated SOC ability was able to produce several correct and usable explanations. This shows that explanatory competence is a distinct analytic skill and not a byproduct of technical experience or classification accuracy.

\vspace{1mm}

\noindent
\textbf{Information Availability vs. Accessibility.} A frequently mentioned source of uncertainty (in the post-study questionnaire) was ``missing information.'' However, nearly all information participants requested, which often includes system roles, IP ownership, and port information, was fully available directly in the logs or in the asset database. What seemed to be missing was not information, but the analysts' ability to find, filter, and process such information under time pressure. This indicated that analysts often struggle with dealing with large volumes of data. 
Hence, we conjecture that wrong or missing explanations are the result of difficulties in structuring information rather than from a lack of data.

\vspace{-2mm}

\begin{cooltextbox}
    \textsc{\textbf{Answer to RQ1:}}
    Our results show a disconnect between analysts’ ability to classify alerts correctly vs. their ability to provide correct and evidence-based explanations for their decisions. Correct classifications were guided by intuition rather than by the identification and articulation of the technical indicators. 
\end{cooltextbox}

\vspace{-2mm}

\subsection{Differential Difficulty Across False Positive Types (RQ2)} 
\label{ssec:rq2}

\noindent 
Our second RQ investigates whether certain types of false positives are more difficult to evaluate and what characteristics make these cases challenging. 

\vspace{1mm}

\noindent
\textbf{Contextual Differences in Performance.} We refer to Table~\ref{tab:participant_results}, showing the classification accuracy per case (``C'' column). The results show substantial variations. For instance, all twelve participants correctly classified \textit{Case 1}, while only nine classified \textit{Cases-3-4-5} correctly. The differences can hardly be explained via randomness:\footnote{A binomial stat test reveals ($p\!<\!.05$) that participants are not answering randomly. Moreover, comparing, e.g., \textit{Case 1} and \textit{Case 3} with a McNemar’s test shows that if the two cases were equally difficult, then results at least as imbalanced as observed would occur \scmath{\approx}8\% of the time. The same holds for, e.g., \textit{Case 1} and \textit{Case 4} (or \textit{5}).} each case is semantically different (see Table~\ref{tab:cases}). Cases can be grouped according to structural features: 
{\footnotesize \textit{(a)}}~clear false (\textit{Case 1}) and true (\textit{Case 6}) positives involved distinct signatures (e.g., known scanner identity and malware quarantined by EDR), which required minimal contextual inference; 
{\footnotesize \textit{(b)}}~context-dependent false positives (\textit{Case~2}) depended on internal business logic not encoded in the log data, and required further investigation or knowledge to be properly explained 
{\footnotesize \textit{(c)}}~ambiguous and mixed-signal false positives (\textit{Case~4}, \textit{Case~5}) required cross-field correlation and interpretation of indicators distributed across the dataset, which makes them the hardest to explain and least reliably classified; 
whereas {\footnotesize \textit{(d)}}~true positives with distributed indicators (\textit{Case~3}) required participants to correlate information across multiple log fields to reconstruct the blocked C\&C communication attempt, wherein classification accuracy was relatively high and explanation quality varied.

\vspace{1mm}

\noindent
\textbf{Structural Characteristics Driving Difficulty.} We identified a set of structural alert features that consistently influenced both classification accuracy and explanation quality: \textbf{(1)}~\textit{Indicator Dispersion}: cases with indicators spread across multiple log fields (e.g., \textit{Case~5}) showed higher cognitive load. Participants had to synthesize information such as timestamps, IP relationships, behavior indicators, and log-source metadata. Several participants failed to understand these signals, which lead either to misclassifications or correct classifications with incorrect explanations; \textbf{(2)}~\textit{High Log Volume}: \textit{Case~4} contained over 31,000 events, which is much larger compared to other cases. While the cause (Windows connectivity tests) was considered benign, the large volume increased perceived complexity. Participants showed uncertainty regarding which events were relevant, which lead to them focusing on irrelevant fields; \textbf{(3)}~\textit{Contextual Ambiguity}: Some false positives required organizational knowledge, which is not part of the alerts themselves. \textit{Case~2} demonstrates that correct classifications were common but explanations were almost always incorrect. This shows that contextual false positives are still difficult to deal with even in cases when technically trivial; and \textbf{(4)}~\textit{Similarity to Attack Patterns}: false positives that represent real attacks (e.g., retransmissions or unusual domain redirects) created hesitation among participants. They frequently misinterpreted harmless behavior as threatening unless an anchor (e.g., scanner identity, benign domain reputation) was obvious. 

\vspace{1mm}

\noindent
\textbf{Cognitive and Organizational Factors Relevant to Difficulty.} Several cognitive and environmental factors also influenced perceived difficulty. Time pressure was the reason the analysis was cut short, which led analysts to fixate on initial indicators. Furthermore, participants stated that tool navigation challenges made it difficult to identify relevant fields even when visible. This was also the case 
with mental models of the network, which led participants to misinterpret IP relationships or misunderstand system roles. Finally, heuristic shortcuts often replaced systematic reasoning, particularly in unclear cases. Note that such feedback was not meant as a critique to our study setup, but rather, to the overarching problem of routinely triaging alerts in their SOC.

\vspace{1mm}

\noindent
\textbf{Cross-Participant Patterns.} The experiences of top and bottom performers further elaborated case difficulty. Top performers recognize case structure early and adjust their strategy accordingly. Bottom performers relied on single-field inspection or their prior experiences. Interestingly, participants with high IT experience struggled with contextual cases. This suggests that the reason behind this difficulty are the properties of the alert and not analysts themselves.

\vspace{-2mm}

\begin{cooltextbox}
    \textsc{\textbf{Answer to RQ2:}}
    Across our six cases, there is a substantial difference in the classification correctness of our participants.
    Our results suggest that triaging certain alarms requires different mental load: some alerts offer clear technical anchors, while others require complex and often contextual interpretation. Triaging became challenging when relevant indicators were not evident (e.g., spread across multiple log sources, or embedded in high-volume data). These characteristics reduced analysts’ ability to classify (and explain) alerts reliably. 
\end{cooltextbox}

\section{Discussion and Implications}
\label{sec:discussion}

\subsection{Lessons Learned and Limitations}
\label{ssec:limitations}

\noindent
\textbf{Lessons Learned.} We identify three major takeaways.
\begin{itemize}
    \item Our study suggests that our analysts’ decision-making processes are shaped based on the provided information rather than by the complexity of its the root causes. Several cases demonstrated that even routine false positives became difficult in cases when contextual cues were scattered, incomplete or presented in a format that required significant manual efforts and reconstruction. On the other hand, technically complex alerts could be handled reliably whenever relevant attributes were ``obvious'' and provided to the analysts in a coherent way. Thus, we argue that the practical bottleneck in SOC triage is not the volume of data, but how the data is organized and presented. 
    
    \item A second lesson learned relates to time management. Most participants reported that the allotted time was not enough (although two said otherwise),
    which impacted their performance. This is a valid observation. Yet, time management is a crucial component in SOCs. In fast-paced environments, such as SOCs, it is imperative to make decisions quickly (and accurately). The fact that some cases required ``more time'' highlights {\footnotesize \textit{(i)}}~that some alarms are harder to triage---supporting RQ2; but also {\footnotesize \textit{(ii)}}~that the way information is provided to the analyst is not optimal. Nonetheless, after receiving feedback, some participants stated they learned new knowledge from this exercise.
        
    \item The third lesson learned stems from the fact that, in our study, no AI-driven assistance was received. More specifically, the considered SOC, at the time of our study, did not have operational AI-powered systems (e.g., RAG or LLMs). Moreover, in the pre-study questionnaire, five participants stated that ``they use AI but feel unsafe doing so'', whereas the remaining seven stated that they do not use AI (although five reported they are unsure of whether AI is useful in their context). We believe that such a divergence underpins an operational dilemma: AI could be used to better organize the information scrutinized by the analyst, but analysts themselves do not (at the time of our study) see AI as a trustworthy assistant---at least for security-sensitive tasks.
\end{itemize}

\noindent
\textbf{Limitations.} We discuss potential critiques that could be made to our research.
\begin{itemize}
    \item \textit{SLR.} The paper-collection phase of our SLR was done by one author who double checked the results, whereas the qualitative analysis was done by two authors to minimise bias and increase correctness. Our search queries should be specific enough to capture all works relevant to answer RQ0, but to avoid missing relevant works we also checked if, among the references in the considered papers, we could find titles that indicated works relevant for RQ0. To the best of our knowledge, our SLR is exhaustive.

    \item \textit{Sample size.} It would be unfair to criticize our study for the ``small sample size''. It is well-known that publicly-available and up-to-date information on how SOCs are run internally, as well as what data flows through their SIEMs and what alarms are triggered, can hardly be found. Indeed, SOC-related research is mostly driven by direct collaboration with operational SOCs~\cite{van2022deepcase,alahmadi202299,kersten2024security}. Such studies are, therefore, particularly challenging to carry out---because finding agreements with professionals is not simple in the general sense~\cite{kaur2022needed,horstmann2025sorry}, and particularly so when the subject of discussion is security-sensitive content. Consequently, user studies in a SOC setting are relatively small scale see our repository~\cite{repository}. In contrast, in other domains (e.g., phishing) valid participants can be regular employees~\cite{acharya2022human} or internet users~\cite{yeke2024wear}, enabling a broader coverage and facilitating generalizations. The size of our sample (12) aligns with the average of extant literature, and our sample is larger than that of the closest work~\cite{kersten2024security} (which has 4). As we wrote (§\ref{ssec:gap}), we do not claim generalizability.

    \item \textit{Selection of Cases.} In designing our study, we were time-constrained. Given that each participant had to partake in a debriefing phase, as well as in the pre-/post-questionnaires (which required $\approx$90 minutes altogether), we could not put dozens of cases. This inevitably forced us to handpick a select number of cases that exhibited enough variety to answer our RQs. We acknowledge that our selection may have been driven by subjectivity. However, given that we are not claiming generalizable results, we do not believe that our choices would threaten the validity of our conclusions. Nonetheless, we provide additional information on selected cases in our repository~\cite{repository}: such information can be useful to future work for designing studies similar to ours.

    \item \textit{Individual setting.} Analysts were required to work without interacting with colleagues, which is a common practice in SOC workflows. We isolated individual reasoning to avoid cross-influence, but this also removes the social mechanisms (e.g., discussion, verification, escalation) that mitigate uncertainty. In regards to the provided explanation, these were deliberately open-ended (to differentiate from~\cite{kersten2024security}). The downside is that correctness assessment dependent on qualitative (mostly inductive) coding. To mitigate this, we adopted a three-party reviewing system (see §\ref{ssec:coding}), which also involved the cooperation of the examined SOC, before drawing our conclusions. 
    
    \item \textit{Trustworthiness.} A tacit assumption is that our participants carried out this exercise faithfully. All participants were properly debriefed on the expectations of our study (and we were available to provide guidance during the exercise), but we had no control on whether participants truly did their best in our field exercise. For ethical reasons, there was no incentive in performing well.
    
\end{itemize}
\vspace{-1mm}
\noindent
Finally, we reiterate that this work was done in cooperation with the examined SOC. This peculiarity substantially increased the ecological validity of our study: first, because of the selection of appropriate \textit{cases} and the overall design of the experiment; second, because of the qualitative evaluation. Indeed, discrepancies among coders were resolved by reviewing the examined SOC's specific context. Importantly, the major findings of our work go against what would be expected of any SOC's desiderata (i.e., the fact that analysts cannot properly explain their decisions can be seen as a weak point), which is why we argue that our results are not biased/affected by potential conflicts of interests.

\subsection{Positioning within Extant Work}
\label{sssec:extant_work}

Prior studies (e.g.,~\cite{alahmadi202299,mink2023everybody,saha2025expert}) have primarily examined analysts’ practices at a conceptual or procedural level (e.g., mapping workflows, documenting cognitive challenges or highlighting organizational constraints). These works provide broad insights into how analysts report making decisions but they do not characterize how well analysts reason about alerts when provided with the concrete technical evidence. Hence, we focus not on perceived obstacles or self-reported strategies, but on the observable reasoning behavior when practitioners engage with real alerts from their operational environment. This shift from descriptive accounts to empirical evaluation allows for better understanding of SOC decision-making.

Earlier evaluations that included participants in hands-on tasks, such as the simulations in~\cite{zhong2017studying,zhong2018learning} or the alert relevance assessments in~\cite{kurogome2019eiger}, focused on whether participants could identify suspicious activity. These studies aim to showcase analysts’ general ability to detect threats. However, they do not examine how analysts formulate their interpretations. In contrast, our work uncovers a remarkable difference between selecting the correct classification and articulating a technically accurate explanation. This distinction matters because SOC workflows depend not only on correct decisions but also on coherent documentation for handovers, incident reports and auditability. These are are all components that, to our knowledge, prior studies did not empirically measure.

Let us compare our findings with those of Kersten et al.~\cite{kersten2024security}, whose field exercise included 10 alerts, each analysed by 4 analysts. Out of 40 responses, 36 were correct (all subjects correctly classified 9 out of 10 alerts). Of the 4 misclassifications, 3 were for ``interesting'' alerts, which implied more complex cases to triage; these results somewhat align with ours. Overall, the median time to investigate an alert was 10 minutes; however, the authors of~\cite{kersten2024security} state that ``It is likely that SOC analysts analyze these incidents faster in the SOC than in our experiment as we employed a think-aloud protocol'', thereby explaining why participants took more time than in our study. We could not find specific details about the correctness of the explanations. However, and crucially, we cite the following excerpt: ``we found that all errors but one were due to differences in the experiment setup from the operational setup the analysts are used to.''

\subsection{Recommendations}
\label{sssec:recommendations}

The topmost pain-point mentioned by our participants refers to the way information is presented. Hence a recommendation is to \textbf{improve data presentation/visualization} (e.g.,~\cite{ulmer2019netcapvis}). This can be achieved by consolidating key indicators, contextual attributes and historical data within SOC platforms.
Such a suggestion can also encompass automated enrichment processes (e.g., embedding asset metadata, known business processes or recent infrastructure changes directly into the alert) as well as explainable-AI techniques (e.g.,~\cite{van2022deepcase}). 

Furthermore, \textbf{we recommend SOC engage in activities similar to those in our work}. The poor performance of our participants may also stem from a lack of experience in explaining decisions in an open-text format. It is possible that analysts knew the reason why an alert was (or not) an FP, but they formulated their thoughts poorly. More frequent ``explanation-based'' exercises could reinforce the analysts' abilities to communicate their operational decisions.

Finally, the difficulty of triaging different alarms can vary significantly. As a result, SOC improvement efforts should not treat all alerts equally. Instead, alerts that repeatedly caused uncertainty (e.g., those requiring cross-system correlations or business knowledge) should be prioritized. This could include \textbf{adjusting SIEM rules, improving documentation, or integrating clearer contextual indicators}. Moreover, an argument can also be made on the importance of \textit{periodic review sessions}, which can be used to retroactively identify the most common ``trivial'' or ``relevant'' alerts are encountered by analyst.

\vspace{-2mm}

\section{Conclusions}
\label{sec:conclusions}

Our real-world field study with 12 SOC analysts shows that while analysts generally make correct triage decisions, they often struggle to explain them. We used six operational alarm cases drawn directly from the SOC’s production environment, wherein we observed that 83\% of classifications were correct. However, only fewer than half of the explanations aligned with the true root cause. Some of the alarms were harder to reason about, regardless that analysts had a full access to the standard SIEM, case-management tools and asset context. This indicates that the core challenge lies in interpreting and articulating context rather than the availability of data. Our findings expose a practical risk for SOC operations, where transparent decision-making is of utmost importance. 

Future work can use our experimental approach and findings to carry out similar studies in similar (or different) SOCs. We cannot claim that the analysts of all SOCs are unable to provide the right justification. Yet, our SLR showed that limited attention has been given to such a pivotal aspect of a SOC's workflow. More research is needed to further explore this operational dimension.

\vspace{1mm}
\noindent
\textbf{Acknowledgment.} We thank the participants of our field study for the time they invested in this effort. We also extend our thanks to the DIMVA PC for the great feedback. Parts of this research was funded by Hilti.


\bibliographystyle{splncs04}


\begin{thebibliography}{10}
\bibitem{sopan2018building}
Sopan, A., Berninger, M., Mulakaluri, M., Katakam, R.: Building a machine learning model for the soc, by the input from the soc, and analyzing it for the soc. In: IEEE VizSec (2018)
\end{thebibliography}

\appendix

\section{Additional Details}
\label{app:details}

\subsection{Annex of the Systematic Literature Review}
\label{sapp:slr}

\noindent
The preliminary check to get an initial understanding of the state of the art was done qualitatively by two authors in Dec. 2024. At the time, we could not find any work that focused on our RQs, which led to us carrying out our study. 

The paper collection phase of our SLR was done (in Dec. 2025, prior to our submission to DIMVA'26) procedurally by entering the queries and retrieving the results provided by Google Scholar. For defining the search terms, they stem from the phrasing of RQ0. Specifically: 
``\textit{SOC}/\textit{alert}/\textit{explainability}''-related keywords are straightforward given our focus; whereas those related to ``user study'' have been inspired by the SLR described in~\cite{pekaric2025we}, which focused on analysing papers accepted to HICSS as either ``technical'' or ``user study'' (or both), and specifically used the term ``survey'' as a subgroup of user studies. To exclude non-peer-reviewed works, we checked if, for any given result, there was a corresponding entry in a peer-reviewed venue. These procedures do not involve any form of qualitative assessment, but were repeated twice (both in Dec. 2025).

For analysing the papers included in our SLR, we adopted a dual-reviewer system (similarly to~\cite{schroer2025sok,pekaric2025we}): two authors reviewed each work and then discussed their findings to reach a consensus. The focus was on ascertaining 
{\footnotesize \textit{(a)}}~if the paper carried out an \textit{user study} and, if so: 
{\footnotesize \textit{(b)}}~\textit{who participated} in the user study---both in terms of sample size and participant's background, e.g., SOC practitioners, or other experts; as well as 
{\footnotesize \textit{(c)}}~the \textit{type} of user study---e.g., a survey, an interview, or a field experiment (like the one we did in our paper); and, finally, {\footnotesize \textit{(d)}} if the paper had strong ties with SOC-related contexts. The analyses of two authors were compared and meetings were held to reach a consensus in cases of discrepancies (we also had one meeting to validate our final results in February 2026). Such an approach increases the reliability our findings~\cite{franklin2001reliability}.

\subsection{Annex to the Ethical Considerations}
\label{sapp:ethics}

\noindent
Our institutions do not mandate an explicit Ethic Review Board (ERB) approval to carry out the research discussed in this work; moreover, at the time of carrying out our study, no formal ERB process existed at our institutions. Nevertheless, we followed established ethical guidelines in our study~\cite{bailey2012menlo}.

First, participants were made aware of the nature of our study, and that their responses would be used for research purposes. Such ``debriefing and request for consent'' was carried out in two phases: upon sending the invitation to participate in the field experiment; and at the beginning of the field experiment. 

Second, our study complied with the ethical standards of the partnered SOC/company. Importantly, it was ensured that participating in the study would have no negative impact on the participants. In addition, no psychological pressure was exerted on the participants, and the fact that the field study was not of a professionally-evaluative nature had been clearly communicated. 

Third, all data was collected in an anonymous fashion---and the participants were made aware of this. Furthermore, to preserve our participants' anonymity, we only release aggregated data or (short excerpts of) ``translated'' data, meaning that it is virtually impossible to link any of the information reported in this paper (or any of its derivative) to the specific individual. Even when sharing data among the researchers involved in this paper, we ensured that such data was exchanged using systems/software linked to our institutions.
\subsection{Sample Description (aggregated, for anonimity/confidentiality)}
\label{sapp:demographics}

\noindent
We report additional information obtained via our preliminary questionnaire.

\textbf{Demographics.}
12 SOC employees participated in our study. Six have an MSc. degree, one a BSc., and five have completed an apprenticeship. One has a school-leaving certificate, while another dropped out of university. 11 participants completed their training in the IT sector, and one was specifically trained in IT security.
In terms of professional experience, $>$70\% have 10+ years of experience in the IT sector. One person has $<$1 year of experience, while the remaining participants have 3--10 years of professional experience.
As SOC analysts, 4 have been working for $<$6 months, while the rest have been in this role for 6--36 months. All participants had previously worked in other areas of IT. Network engineering (65\%), system administration (57\%), software development (50\%), and IT helpdesk (42\%) were mentioned particularly frequently.

\textbf{Self-perception of competences} Seven people stated that they have basic knowledge. Four people categorised themselves as advanced in a specific area. One person rated themselves as advanced in several areas. 

\textbf{Technological background}
In terms of the tools used, 85\% stated that they work regularly with a SIEM tool. 65\% already had experience with packet capture technologies and 50\% with firewalls. 28\% of the participants had already worked with threat intelligence, while intrusion detection systems (IDS) and detection \& response (EDR) were mentioned less frequently (resp. 14\%, 20\%). 

\textbf{Log analysis experience.}
92\% stated that they had experience in analysing logs. Windows logs were mentioned most frequently (65\%), followed by firewall logs (50\%), web server and proxy logs (43\% each), and EDR (8\%) logs.

\textbf{Assessment of alarms.}
Participants were asked to rate their confidence in {\footnotesize \textit{(a)}}~interpreting and {\footnotesize \textit{(b)}}~recognising security alarms on a 1--5 scale. For the former, the average score was 2.64; for the latter, the mean value was 2.71. 

\textbf{Dealing with uncertainties}
92\% stated that they consulted colleagues in case of uncertainties; however 71\% participants also relied on their own research, analysing additional logs and comparing with existing playbooks. Only 33\% stated that they would create a ticket and pass the case on if they were unsure.

\textbf{Attitude towards use of AI.}
Regarding AI support, two participants stated that they do not use AI and feel safe doing so. The remaining ten people were divided into two groups: Half use AI but feel unsafe doing so. The other half do not currently use AI, but are also unsure whether it would be useful.
\vspace{-1mm}
\textbox{\textbf{Remark.} Some of our participants had little expertise with some tools. However, this simply denotes that such participants do not use such tools in their daily routines. This does not change the fact that such people are professional SOC employees, and that such tools are always available to be consulted for SOC-related duties. Hence, the ecological validity of our study is preserved.}
\subsection{Exemplary codes}
\label{sapp:examples}
We provide some examples of our reasoning for \textit{Cases 1} and \textit{2}. For \textit{Case 1}, all coders agreed that a correct explanation had to mention ``nessus scan'' or equivalent; an exemplary incorrect explanation stated ``client access to the internet'', deemed vague by all coders. For \textit{Case 2}, only P12 provided a valid explanation with ``developer test'': despite being short, it showed that P12 understood that it was an FP because it was due to some explicitly allowed internal tests triggering alarms; in contrast, a wrong explanation by P6 was ``download Ubuntu'': while not strictly incorrect, such software-download operations \textit{should be treated as TPs in this SOC as they can conceal malicious activities and demand further scrutiny}. (\textit{Case 2} is an FP is because there was an explicit ``developer rule'' that was triggered, so the case should have been closed as an FP (and only P12 made it clear). Intriguingly, R3 considered P12's answer as \xmark{} (because it was vague to R3), but R1 deemed it phrased in a contextually-correct way.

\end{document}